\documentclass{article}
\usepackage{spconf,amsmath,graphicx}
\usepackage{amssymb}
\usepackage{enumitem}

\title{Adversarial Stimuli:\\
Attacking Brain-Computer Interfaces via Perturbed Sensory Events}
\name{Bibek Upadhayay \& Vahid Behzadan}
\address{SAIL Lab\\
University of New Haven, West Haven, CT - 06516}
\begin{document}

\maketitle

\begin{abstract}
Machine learning models are known to be vulnerable to adversarial perturbations in the input domain, causing incorrect predictions. Inspired by this phenomenon, we explore the feasibility of manipulating EEG-based Motor Imagery (MI) Brain Computer Interfaces (BCIs) via perturbations in sensory stimuli. Similar to adversarial examples, these \emph{adversarial stimuli} aim to exploit the limitations of the integrated brain-sensor-processing components of the BCI system in handling shifts in participants' response to changes in sensory stimuli. This paper proposes adversarial stimuli as an attack vector against BCIs, and reports the findings of preliminary experiments on the impact of visual adversarial stimuli on the integrity of EEG-based MI BCIs. Our findings suggest that minor adversarial stimuli can significantly deteriorate the performance of MI BCIs across all participants (p=0.0003). Additionally, our results indicate that such attacks are more effective in conditions with induced stress.
\end{abstract}
\section{Introduction}
\label{sec:Intro}
In recent years, Brain-Computer Interfaces (BCIs) have enjoyed accelerating rises in accessibility and development. These systems provide a direct medium between the brain and computing devices, thereby facilitating a plethora of applications, ranging from medical monitoring and rehabilitation \cite{wheelchair,robotichand} to enabling seamless modes of communications and control. 
The foundational mechanisms of such BCI applications are based on exploiting the correlations between measurable electrochemical activities of the brain and the higher-order cognitive functions such as perception and motor intent \cite{kosmyna2018attending}. While the variety of measurement techniques adopted in BCIs has been increasing, ElectroEncephaloGraphy (EEG) has been dominating the field due to its relative simplicity and accessibility \cite{ekandem2012evaluating}. The increasing availability of commercial EEG devices has facilitated the rise of numerous novel applications, from brain-controlled keyboards \cite{chaudhary2022spelling}, neuroprosthetics \cite{robotichand}, and controlling wheelchairs \cite{wheelchair} to game play \cite{tangermann2008playing} and biometric authentication \cite{nguyen2013motor} 

However, such recent advances in BCI technologies give rise to new concerns about the security of these solutions. In response, the field of \emph{Neurosecurity} \cite{denning2009neurosecurity} has been established to address the security and privacy issues that may arise from the accelerating adoption of BCI technologies. The growing body of work on this field includes studies on the vulnerabilities of the hardware, software, and machine learning components of BCI systems \cite{bernal2021StateofArtsecurity}. Instances of the latter are the studies that propose Man-in-The-Middle (MiTM) attacks, in which the adversary compromises the connection between the EEG headset and the computer, and injects adversarially crafted noise to manipulate the machine learning model processing the correlations between measurements and intent \cite{landau2020mind,li2015brain,vadlamani2016jamming,frank2017using,zhang2019vulnerability,liu2021universal}. However, current studies fail to present attacks which exploit vulnerabilities that are specific to BCIs.

In this paper, we aim to address the aforementioned shortcoming by exploring the feasibility of manipulating Motor Imagery (MI) BCIs via perturbing the visual stimuli observed by the BCI user. Inspired by adversarial example attacks against machine learning, we hypothesize that the integration of cognitive, measurement, and machine learning components in EEG-based BCIs may also be vulnerable to minor perturbations that can be induced directly at the sensory level. To examine this hypothesis, we performed a preliminary study on 7 human subjects to measure the impact of visual adversarial stimuli on their performance in MI tasks. The results of these experiments validate the feasibility of adversarial stimuli attacks against EEG-based MI BCIs.

The remainder of this paper is organized as follows: Section \ref{sec:adv_stimuli} introduces adversarial stimuli as an attack vector against BCIs. To establish the feasibility of this attack vector, Section \ref{sec:experiment} presents the details of preliminary experiments on proof-of-concept adversary stimuli attacks against EEG-based MI BCI tasks. Section \ref{sec:results} reports the results of these experiments, and Section \ref{sec:conclusion} concludes the paper with a discussion on our findings, as well as remarks on future directions of research.

\section{ADVERSARIAL STIMULI} 
\label{sec:adv_stimuli}
Previous studies have established that the machine learning models trained on EEG-based MI tasks do not generalize well and are prone to overfitting \cite{xu2020cross,mousavi2021improving}. Therefore, similar to all machine learning models, such BCI models are also prone to adversarial examples attacks \cite{zhang2019vulnerability,liu2021universal,jiang2019active}. These findings raise the hypothesis that the EEG-based BCIs may also be vulnerable to perturbations in the sensory domain. 

Accordingly, we define \emph{Adversarial Stimuli} as perturbations in sensory events introduced by adversaries with the intent of tampering with the BCI performance. These stimuli can be in the form of auditory, visual, or tactile perturbations in the environment surrounding the targeted BCI user. In the case of visual adversarial stimuli, an example of such perturbations can manifest in the form of irregular flickering added to the scene observed by the user. This flickering effect can be implemented in either the entire observable scene (e.g., flickering in the light sources such as lamps), or may be contained to specific regions or observable events (e.g., a segment of the visual interface in the BCI ecosystem). Consequently, such perturbations do not require the attacker to have local access to the environment, and can be induced completely remotely.

\section{EXPERIMENTAL SETUP} 
\label{sec:experiment}

To investigate the feasibility of adversarial stimuli attacks, we formulate the following hypotheses: \emph{Hypothesis 1 (H1): } There are small perturbations in the visual observations that can negatively impact the performance of the EEG-based MI BCIs; and \emph{Hypothesis 2 (H2):} The effects of such adversarial attacks are more significant in the stressful settings of time-constrained tasks than on the open-time task.

We examined these hypotheses via experiments with seven subjects comprised of 2 males and 5 females aged 23-31. The subjects were asked to use an MI BCI to play the classic game of Pong displayed on a computer screen. We used the Emotiv Epoc X EEG headset for these experiments with 14 channels (AF3, F7, F3, FC5, T7, P7, O1, O2, P8, T8, FC6, F4, F8, AF4) with a sequential sampling rate of 128 Hz. The EmotivBCI software was used to create individual user profiles with individual training profiles. 

 In order to familiarize our participants with the BCI settings, the participants were first instructed to train an MI model for the task of moving a cube up or down inside the default environment of EmotivBCI. To facilitate uniformity, the participants were asked to imagine the act of \emph{throwing a ball with their right hand} to initiate the MI signal for the ``up'' action. Similarly, in order move the cube in ``down'' direction, the participants were asked to imagine \emph{kicking a ball using their right leg}. 

After successful training, we evaluated the trained model by asking the participants to follow a random sequence of instructions (i.e., directions) in the cube task. We continued with the experiment only if the participants succeeded in implementing the instructed actions. Otherwise, we restarted the training process. 

For the subsequent stages of our experiments, we created two customized versions of Pong, namely: the \emph{Warm-up} environment and the \emph{full game}. Each of these environments could be configured to operate in either the \emph{normal} or \emph{adversarial} modes. The warm-up environment provides a simplified setting to practice the MI control of the paddle by eliminating the ball movement - that is, the ball remains stationary. The environment provides instructions to participants to move the paddle in up or down directions, and the participants succeed if they manage to move the paddle in instructed direction within 15 seconds. Each participant could score up to 12 points in 3 cycles. 

\begin{figure}
\centering
    \includegraphics[width=0.5\textwidth]{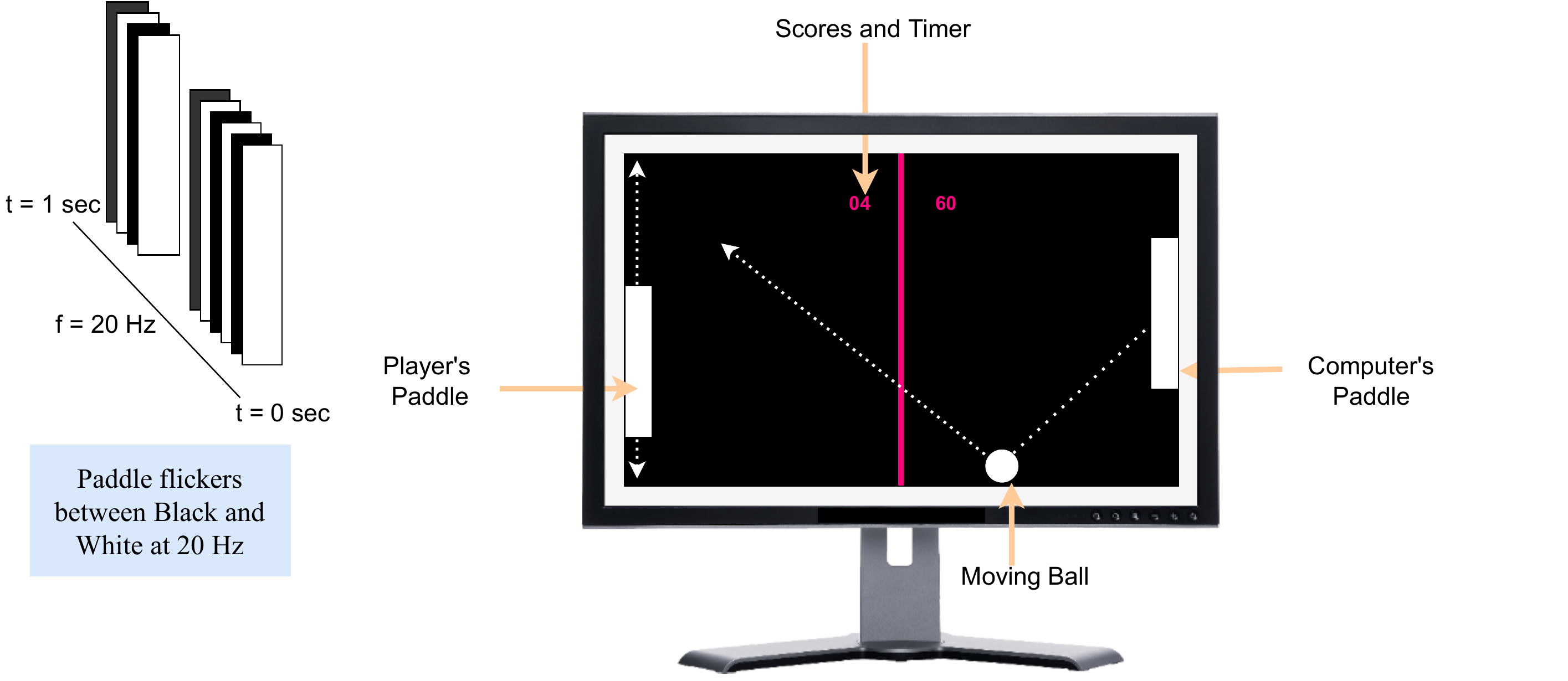}
    \caption{\emph{Full game} Environment (Env 3 and Env 4)}
    \label{fig:time_stress_env}
    
\end{figure}

In the full game settings, the participants play the Pong game against an automated player by moving the paddle up or down to hit the ball, and score a point for each time they succeed. The game ends when the participants miss more than 5 balls, or the game duration exceeds 3 minutes. Fig. \ref{fig:time_stress_env} illustrates this environment.

In the adversarial modes, we simulate a scenario in which an adversary has gained access to the Pong environment and can at any time change the flickering rate of the paddle and the ball. In our experiments, the adversarial flickering rate was set to 20 Hz. In the normal environment there is no perturbation by the adversarial stimuli, whereas in the adversarial environment the perturbation is added at random intervals.

\section{RESULTS}
\label{sec:results}

For each of the \emph{warm-up} and \emph{full game} environments, we measured the total score obtained by participants under both normal and adversarial modes. Furthermore, we also measured the error rate for each subject, $Er = (S_n - S_a)/S_n$, where $S_n$ is the score achieved in the normal mode, and $S_a)$ is the score obtained in the adversarial mode. For experiments in the warm-up mode under normal mode (E1) and adversarial mode (E2), the maximum achievable score was set to 12. In the \emph{full game} experiments under normal mode (E3) and adversarial mode (E4), no hard limit was set on the maximum score.

In order to test both of our hypotheses, we composed the following null hypotheses: Null hypothesis I states that there is no significant difference in the performance between normal mode and adversarial mode in the \emph{warm-up} environments; and Null Hypothesis II states that there is no difference between the impact of adversarial stimuli in \emph{warm-up} environment and \emph{full game} environment. We performed paired t-tests with significance level $\alpha = 0.05$  for each case, namely E1 vs. E2 and E3 vs. E4. We also performed a paired t-test for the overall normal vs.  adversarial environments combining \emph{warm-up} and \emph{full game} (i.e., E1, E3 vs.  E2, E4). The statistical test results are presented in Table \ref{table:stat_test_hypo-1}. We observed that the value of $t \gg t_{c}$, and $p\ll 0.05$ in all of the aforementioned cases, thus rejecting the null hypotheses. The rejection of the first null hypothesis supports the claim that \textbf{there is significant deterioration of performance between the normal mode and adversarial mode in the \emph{Warm-up} environment.} With regards to Hypothesis II,  our statistical paired t-test with the significance level $\alpha = 0.05$ and a critical value for a right-tailed test if $t_{c} = -1.943$, result in the value of $t= -6.04 \ll t_{c}$, and $p= 0.00005 \ll 0.05$, hence we concluded to reject the null hypothesis. The statistical tests signify that there is a significant increase in the error rate in \emph{full game} conditions than in \emph{Warm-up} environments when attacked via adversarial stimuli. Hence we accept our alternate hypothesis that \textbf{that the impact of the adversarial attack are more significant on the time-constrained task than on the warm-up task.}

\begin{table}[]
\centering
\begin{tabular}{ccccc}
\cline{2-5}
\multicolumn{1}{c|}{}         & \multicolumn{2}{c|}{\emph{Warm-up} Envs}                      & \multicolumn{2}{c|}{\emph{Full game} Envs}                          \\ \cline{2-5} 
\multicolumn{1}{c|}{}         & \multicolumn{1}{c|}{Normal} & \multicolumn{1}{c|}{Adversarial} & \multicolumn{1}{c|}{Normal} & \multicolumn{1}{c|}{Adversarial} \\ \hline
\multicolumn{1}{|c|}{Subject} & \multicolumn{1}{c|}{Env 1}  & \multicolumn{1}{c|}{Env 2}       & \multicolumn{1}{c|}{Env 3}  & \multicolumn{1}{c|}{Env 4}       \\ \hline
\multicolumn{1}{|c|}{S1}      & \multicolumn{1}{c|}{7}      & \multicolumn{1}{c|}{5}           & \multicolumn{1}{c|}{15}     & \multicolumn{1}{c|}{8}           \\ \hline
\multicolumn{1}{|c|}{S2}      & \multicolumn{1}{c|}{5}      & \multicolumn{1}{c|}{3}           & \multicolumn{1}{c|}{16}     & \multicolumn{1}{c|}{6}           \\ \hline
\multicolumn{1}{|c|}{S3}      & \multicolumn{1}{c|}{5}      & \multicolumn{1}{c|}{3}           & \multicolumn{1}{c|}{14}     & \multicolumn{1}{c|}{10}          \\ \hline
\multicolumn{1}{|c|}{S4}      & \multicolumn{1}{c|}{7}      & \multicolumn{1}{c|}{4}           & \multicolumn{1}{c|}{24}     & \multicolumn{1}{c|}{7}           \\ \hline
\multicolumn{1}{|c|}{S5}      & \multicolumn{1}{c|}{7}      & \multicolumn{1}{c|}{4}           & \multicolumn{1}{c|}{17}     & \multicolumn{1}{c|}{9}           \\ \hline
\multicolumn{1}{|c|}{S6}      & \multicolumn{1}{c|}{6}      & \multicolumn{1}{c|}{3}           & \multicolumn{1}{c|}{25}     & \multicolumn{1}{c|}{11}          \\ \hline
\multicolumn{1}{|c|}{S7}      & \multicolumn{1}{c|}{7}      & \multicolumn{1}{c|}{4}           & \multicolumn{1}{c|}{9}      & \multicolumn{1}{c|}{5}           \\ \hline
\multicolumn{1}{|c|}{Average} & \multicolumn{1}{c|}{6.28}   & \multicolumn{1}{c|}{3.17}        & \multicolumn{1}{c|}{17.14}  & \multicolumn{1}{c|}{8}           \\ \hline
\multicolumn{1}{l}{}          & \multicolumn{1}{l}{}        & \multicolumn{1}{l}{}             & \multicolumn{1}{l}{}        & \multicolumn{1}{l}{}            
\end{tabular}
\caption{Experiments Results}
\label{table:experiment_results_score}
\end{table}

\begin{table}[]
\begin{tabular}{lrrl}
\cline{2-4}
\multicolumn{1}{l|}{Paired-t-test}                     & \multicolumn{1}{l|}{tc}    & \multicolumn{1}{l|}{t}      & \multicolumn{1}{l|}{p}      \\ \hline
\multicolumn{1}{|l|}{Warm-up Envs (E1 vs.  E2)}          & \multicolumn{1}{r|}{1.943} & \multicolumn{1}{r|}{12.728} & \multicolumn{1}{l|}{0.0001} \\ \hline
\multicolumn{1}{|l|}{Full game Envs (E3 vs.  E4 )}       & \multicolumn{1}{r|}{1.943} & \multicolumn{1}{r|}{4.923}  & \multicolumn{1}{r|}{0.0013} \\ \hline
\multicolumn{1}{|l|}{Normal vs.  Adv (E1, E3 vs.   E2,E4)} & \multicolumn{1}{r|}{1.711} & \multicolumn{1}{r|}{4.58}   & \multicolumn{1}{r|}{0.0003} \\ \hline
                                                       & \multicolumn{1}{l}{}       & \multicolumn{1}{l}{}        &                            
\end{tabular}
\caption{Statistical Test for Hypothesis-I}
\label{table:stat_test_hypo-1}
\end{table}

\begin{table}[]
\centering
\begin{tabular}{lrr}
\cline{2-3}
\multicolumn{1}{l|}{}         & \multicolumn{1}{l|}{ \emph{Warm-up} Envs}        & \multicolumn{1}{l|}{\emph{Full game} Envs}           \\ \hline
\multicolumn{1}{|l|}{Subject} & \multicolumn{1}{l|}{Error Rate (Env 1 to Env 2)} & \multicolumn{1}{l|}{Error Rate(Env 3 to Env 4)} \\ \hline
\multicolumn{1}{|l|}{S1}      & \multicolumn{1}{r|}{0.16}                        & \multicolumn{1}{r|}{0.46}                       \\ \hline
\multicolumn{1}{|l|}{S2}      & \multicolumn{1}{r|}{0.16}                        & \multicolumn{1}{r|}{0.62}                       \\ \hline
\multicolumn{1}{|l|}{S3}      & \multicolumn{1}{r|}{0.16}                        & \multicolumn{1}{r   |}{0.28}                       \\ \hline
\multicolumn{1}{|l|}{S4}      & \multicolumn{1}{r|}{0.25}                        & \multicolumn{1}{r|}{0.7}                        \\ \hline
\multicolumn{1}{|l|}{S5}      & \multicolumn{1}{r|}{0.25}                        & \multicolumn{1}{r|}{0.47}                       \\ \hline
\multicolumn{1}{|l|}{S6}      & \multicolumn{1}{r|}{0.25}                        & \multicolumn{1}{r|}{0.56}                       \\ \hline
\multicolumn{1}{|l|}{S7}      & \multicolumn{1}{r|}{0.25}                        & \multicolumn{1}{r|}{0.44}                       \\ \hline
                              & \multicolumn{1}{l}{}                             & \multicolumn{1}{l}{}                           
\end{tabular}
\caption{Error rate in  \emph{Warm-up} vs.  \emph{Full-game} Environments}
\label{table:experiment_results_score_h2}
\end{table}

\section{DISCUSSION}
\label{sec:discussions}

\textbf{Impact: }In our experiments, we observed that adversarial stimuli can significantly deteriorate the performance of MI tasks in EEG-based BCIs. We also observed that the impact of visual adversarial stimulus is more pronounced in the time-constrained settings, as compared to the warm-up task. This provides evidence that the proposed attack vector can be used by adversaries to manipulate the performance of EEG-based MI applications. Furthermore, the attack by the adversarial stimuli may affect the neural data generation stage that might affect the later stimuli generation \cite{LANDAU2020105932}, resulting in a prolonged effect on BCIs performance. The practical implications of such attacks are significant - Malicious actors can use adversarial stimuli to target the integrity of systems such as MI-based wheelchairs and neuroprosthetics. Similarly, adversarial stimuli can be used for denial of service against EEG-based authentication systems, thereby blocking access to authorized users due to induced authorization failures. We also hypothesize that such attacks are not limited to MI tasks; for instance, changing the flickering frequency of the visual stimuli may affect the performance of SSVEP and VI-based BCIs. We therefore believe that the threat posed by adversarial stimuli warrants further studies on the underlying dynamics and mitigation of such attacks.

\begin{figure}
\centering

    \includegraphics[width=0.45\textwidth]{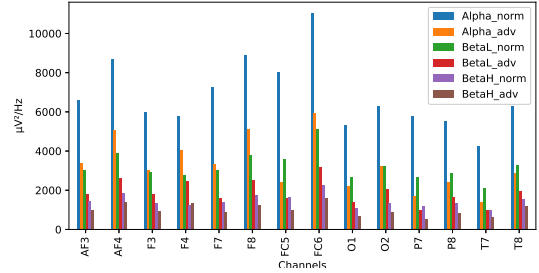}
    \caption{Average power for Alpha(8-12 Hz), BetaL(12-16 Hz) and BetaH(16-25 Hz) bands in normal and adversarial settings.}
    \label{fig:powerband}
    
\end{figure}

\textbf{Preliminary Analysis of Causes:} As an initial step towards investigating the dynamics of adversarial stimuli attacks, we further investigated the \emph{warm-up} environment where the state of the paddle and the intent of the subjects were recorded at each timestep. We calculated the beta-low to alpha and beta-high to alpha ratios in the recorded EEG signals, and observed that both ratios increase from normal to adversarial settings. The increase in the beta/alpha ratio suggests high engagement in cognitive power. The alpha synchronization is considered an important component in the selective attention process where it inhibits the unattended positions during visual spatial orienting \cite{rihs2007mechanisms}. However, we observed that the overall alpha and beta power bands decrease from normal to adversarial settings. The flickering stimulus should have elicited its own frequency or frequencies in its multiples (e.g., 20, 40, 60 Hz) however, we did not observe the rise in power band in those frequencies in the adversarial settings. Additionally, we observe that the adversarial stimulus suppresses the amplitude of both the alpha and beta power. The suppression is significantly more in the alpha band than in any other band as depicted in Fig \ref{fig:powerband}. Furthermore, we observed that in the presence of the adversarial stimulus, motor imagery signals based on mu rhythm (8-12 Hz) are also suppressed, which might be a contributing factor for the deterioration of the overall performance. This observation raises a further question on whether or not the MI signals can be fully disentangled from the SSVEP signal in EEG measurements.

Also, the participants in our experiments were focused on the given task and the perturbation was implemented in a smaller region of the screen only. Even though, the participants were asked only to focus on maximizing their score, we can observe that the score decreases significantly in the adversarial environment. One probable explanation for the performance deterioration could be the divided attention. However, our results demonstrate that the visual selective attention based on the beta-to-alpha ratio is higher in the adversarial settings. Hence we did not find sufficient support for this hypothesis. Another probable explanation could be the prediction error. It has been reported that increases in the alpha-to-beta and gamma-band activities are the reflections of predictive coding in visual processing \cite{strube2021alpha}, which increases attention to irrelevant cues \cite{torrents2021rise}. However, in our case we only observed the decrease in the power bands in adversarial settings.

\textbf{Open Questions: }We observe many similarities between the adversarial stimuli attacks in our experiments and adversarial example attacks against machine learning models. In deep reinforcement learning, it is shown that malicious actors can manipulate the action policies of agents via perturbing the the agent's observations via adversarial examples \cite{behzadan2017vulnerability}. Adversarial stimuli attacks follow a similar process, in which perturbing the perception results in changes in the actions of the integrated BCIs system. This analogy gives rise to further questions: are there optimal or efficient adversarial stimuli, similar to those crafted by adversarial examples generation algorithms \cite{papernot2018sok}, that can effectively and persistently induce incorrect actions in BCIs? Moreover, it has been established that training machine learning models on adversarial examples improves their robustness to adversarial examples. We therefore ask whether similar adversarial training procedures in BCI systems and users could enhance robustness against adversarial stimuli? A further line of inquiry arises from the the observed suppression of the mu rhythm under adversarial stimuli: are MI and SSVEP signals separable, or are these sources fundamentally interrelated? And perhaps the most significant question in this direction is about the source of vulnerability to adversarial stimuli: is it the root cause of this vulnerability in the BCI device and software, or does it stem from inherent limitations of the human cognitive system?

\section{CONCLUSION}
\label{sec:conclusion}
We introduced adversarial stimuli as an attack vector against MI-based BCIs. To demonstrate the feasibility of these attacks, we performed experiments on human subjects playing a video game via MI through a EEG device, and observed that minor and random flickers in the visual observations of the game results in significant deterioration of their performance in the game. We also reported our initial analyses on the possible causes of such vulnerability. Furthermore, we enumerated major directions of inquiry and open questions about adversarial stimuli, with the aim of motivating further research in this area.

\bibliographystyle{IEEEbib}
\bibliography{refs}

\begin{thebibliography}{10}

\bibitem{wheelchair}
I.~Iturrate, J.~Antelis, and J.~Minguez,
\newblock ``Synchronous eeg brain-actuated wheelchair with automated
  navigation,''
\newblock in {\em 2009 IEEE International Conference on Robotics and
  Automation}, 2009, pp. 2318--2325.

\bibitem{robotichand}
Dennis~J McFarland and Jonathan~R Wolpaw,
\newblock ``Brain-computer interface operation of robotic and prosthetic
  devices,''
\newblock {\em Computer}, vol. 41, no. 10, pp. 52--56, 2008.

\bibitem{kosmyna2018attending}
Nataliya Kosmyna, Jussi~T Lindgren, and Anatole L{\'e}cuyer,
\newblock ``Attending to visual stimuli versus performing visual imagery as a
  control strategy for eeg-based brain-computer interfaces,''
\newblock {\em Scientific reports}, vol. 8, no. 1, pp. 1--14, 2018.

\bibitem{ekandem2012evaluating}
Joshua~I Ekandem, Timothy~A Davis, Ignacio Alvarez, Melva~T James, and Juan~E
  Gilbert,
\newblock ``Evaluating the ergonomics of bci devices for research and
  experimentation,''
\newblock {\em Ergonomics}, vol. 55, no. 5, pp. 592--598, 2012.

\bibitem{chaudhary2022spelling}
Ujwal Chaudhary, Ioannis Vlachos, Jonas~B Zimmermann, Arnau Espinosa,
  Alessandro Tonin, Andres Jaramillo-Gonzalez, Majid Khalili-Ardali, Helge
  Topka, Jens Lehmberg, Gerhard~M Friehs, et~al.,
\newblock ``Spelling interface using intracortical signals in a completely
  locked-in patient enabled via auditory neurofeedback training,''
\newblock {\em Nature communications}, vol. 13, no. 1, pp. 1--9, 2022.

\bibitem{tangermann2008playing}
Michael Tangermann, Matthias Krauledat, Konrad Grzeska, Max Sagebaum, Benjamin
  Blankertz, Carmen Vidaurre, and Klaus-Robert M{\"u}ller,
\newblock ``Playing pinball with non-invasive bci.,''
\newblock in {\em NIPS}, 2008, pp. 1641--1648.

\bibitem{nguyen2013motor}
Phuoc Nguyen, Dat Tran, Xu~Huang, and Wanli Ma,
\newblock ``Motor imagery eeg-based person verification,''
\newblock in {\em International work-conference on artificial neural networks}.
  Springer, 2013, pp. 430--438.

\bibitem{denning2009neurosecurity}
Tamara Denning, Yoky Matsuoka, and Tadayoshi Kohno,
\newblock ``Neurosecurity: security and privacy for neural devices,''
\newblock {\em Neurosurgical Focus}, vol. 27, no. 1, pp. E7, 2009.

\bibitem{bernal2021StateofArtsecurity}
Sergio~L{\'o}pez Bernal, Alberto~Huertas Celdr{\'a}n, Gregorio~Mart{\'\i}nez
  P{\'e}rez, Michael~Taynnan Barros, and Sasitharan Balasubramaniam,
\newblock ``Security in brain-computer interfaces: State-of-the-art,
  opportunities, and future challenges,''
\newblock {\em ACM Computing Surveys (CSUR)}, vol. 54, no. 1, pp. 1--35, 2021.

\bibitem{landau2020mind}
Ofir Landau, Rami Puzis, and Nir Nissim,
\newblock ``Mind your mind: Eeg-based brain-computer interfaces and their
  security in cyber space,''
\newblock {\em ACM Computing Surveys (CSUR)}, vol. 53, no. 1, pp. 1--38, 2020.

\bibitem{li2015brain}
QianQian Li, Ding Ding, and Mauro Conti,
\newblock ``Brain-computer interface applications: Security and privacy
  challenges,''
\newblock in {\em 2015 IEEE conference on communications and network security
  (CNS)}. IEEE, 2015, pp. 663--666.

\bibitem{vadlamani2016jamming}
Satish Vadlamani, Burak Eksioglu, Hugh Medal, and Apurba Nandi,
\newblock ``Jamming attacks on wireless networks: A taxonomic survey,''
\newblock {\em International Journal of Production Economics}, vol. 172, pp.
  76--94, 2016.

\bibitem{frank2017using}
Mario Frank, Tiffany Hwu, Sakshi Jain, Robert~T Knight, Ivan Martinovic,
  Prateek Mittal, Daniele Perito, Ivo Sluganovic, and Dawn Song,
\newblock ``Using eeg-based bci devices to subliminally probe for private
  information,''
\newblock in {\em Proceedings of the 2017 on Workshop on Privacy in the
  Electronic Society}, 2017, pp. 133--136.

\bibitem{zhang2019vulnerability}
Xiao Zhang and Dongrui Wu,
\newblock ``On the vulnerability of cnn classifiers in eeg-based bcis,''
\newblock {\em IEEE Transactions on Neural Systems and Rehabilitation
  Engineering}, vol. 27, no. 5, pp. 814--825, 2019.

\bibitem{liu2021universal}
Zihan Liu, Lubin Meng, Xiao Zhang, Weili Fang, and Dongrui Wu,
\newblock ``Universal adversarial perturbations for cnn classifiers in
  eeg-based bcis,''
\newblock {\em Journal of Neural Engineering}, vol. 18, no. 4, pp. 0460a4,
  2021.

\bibitem{xu2020cross}
Lichao Xu, Minpeng Xu, Yufeng Ke, Xingwei An, Shuang Liu, and Dong Ming,
\newblock ``Cross-dataset variability problem in eeg decoding with deep
  learning,''
\newblock {\em Frontiers in human neuroscience}, vol. 14, pp. 103, 2020.

\bibitem{mousavi2021improving}
Mahta Mousavi, Eric Lybrand, Shuangquan Feng, Shuai Tang, Rayan Saab, and
  Virginia~R de~Sa,
\newblock ``Improving robustness in motor imagery brain-computer interfaces,''
\newblock in {\em NeurIPS 2021 Workshop on Distribution Shifts: Connecting
  Methods and Applications}, 2021.

\bibitem{jiang2019active}
Xue Jiang, Xiao Zhang, and Dongrui Wu,
\newblock ``Active learning for black-box adversarial attacks in eeg-based
  brain-computer interfaces,''
\newblock in {\em 2019 IEEE Symposium Series on Computational Intelligence
  (SSCI)}. IEEE, 2019, pp. 361--368.

\bibitem{LANDAU2020105932}
Ofir Landau, Aviad Cohen, Shirley Gordon, and Nir Nissim,
\newblock ``Mind your privacy: Privacy leakage through bci applications using
  machine learning methods,''
\newblock {\em Knowledge-Based Systems}, vol. 198, pp. 105932, 2020.

\bibitem{rihs2007mechanisms}
Tonia~A Rihs, Christoph~M Michel, and Gregor Thut,
\newblock ``Mechanisms of selective inhibition in visual spatial attention are
  indexed by $\alpha$-band eeg synchronization,''
\newblock {\em European Journal of Neuroscience}, vol. 25, no. 2, pp. 603--610,
  2007.

\bibitem{strube2021alpha}
Andreas Strube, Michael Rose, Sepideh Fazeli, and Christian B{\"u}chel,
\newblock ``Alpha-to-beta-and gamma-band activity reflect predictive coding in
  affective visual processing,''
\newblock {\em Scientific reports}, vol. 11, no. 1, pp. 1--15, 2021.

\bibitem{torrents2021rise}
David Torrents-Rodas, Stephan Koenig, Metin Uengoer, and Harald Lachnit,
\newblock ``A rise in prediction error increases attention to irrelevant
  cues,''
\newblock {\em Biological Psychology}, vol. 159, pp. 108007, 2021.

\bibitem{behzadan2017vulnerability}
Vahid Behzadan and Arslan Munir,
\newblock ``Vulnerability of deep reinforcement learning to policy induction
  attacks,''
\newblock in {\em International Conference on Machine Learning and Data Mining
  in Pattern Recognition}. Springer, 2017, pp. 262--275.

\bibitem{papernot2018sok}
Nicolas Papernot, Patrick McDaniel, Arunesh Sinha, and Michael~P Wellman,
\newblock ``Sok: Security and privacy in machine learning,''
\newblock in {\em 2018 IEEE European Symposium on Security and Privacy
  (EuroS\&P)}. IEEE, 2018, pp. 399--414.

\end{thebibliography}

\end{document}